.

# Performance analysis of dual-hop optical wireless communication systems over k-distribution turbulence channel with pointing error

Neha Mishra, Sriram Kumar D., and Pranav Kumar Jha

# Performance Analysis of Dual-Hop Optical Wireless Communication Systems over K-distribution Turbulence Channel with Pointing Error

Neha Mishra[1,a)], Sriram Kumar D.[2,b)] and Pranav Kumar Jha[3)]

[1,2,3]Department of Electronics and Communication Engineering,
National Institute of Technology, Tiruchirappalli, Tamil Nadu, India
[a)]nehamishrabe@gmail.com
[b)]srk@nitt.edu

**Abstract.** In this paper, we investigate the performance of the dual-hop free space optical (FSO) communication systems under the effect of strong atmospheric turbulence together with misalignment effects (pointing error). We consider a relay assisted link using decode and forward (DF) relaying protocol between source and destination with the assumption that Channel State Information is available at both transmitting and receiving terminals. The atmospheric turbulence channels are modeled by k-distribution with pointing error impairment. The exact closed form expression is derived for outage probability and bit error rate and illustrated through numerical plots. Further BER results are compared for the different modulation schemes.

## INTRODUCTION

Transmission of optical signals through the atmosphere in a line of sight manner is known as optical wireless communication (OWC). It has gained significant research interest in recent years, however its performance is highly ill-protected to unfavorable atmospheric conditions which occur mainly due to turbulence and pointing errors. The turbulence arises due to the inhomogeneities in temperature and pressure of the atmosphere. This causes a change in refractive index and leads to the fluctuation of amplitude and intensity in the transmitted optical signals. The pointing error occur due to the sway of high rise buildings as a result of weak earthquakes, dynamic wind loads and thermal expansions. This causes misalignment between source and destination transceivers.

To evaluate the impact of atmospheric turbulence various channel model are developed such as gamma-gamma, negative exponential and log normal model. The log normal distribution is found suitable for weak turbulence regimes were as gamma- gamma distribution is suitable for moderate to strong turbulence regimes. Another channel model is a k distribution channel model which is very popularly used for strong turbulence regimes. Further in literature, we study two types of relays i.e. amplify and forward (AF) and decode and forward relay (DF) relay. In this paper, we consider Decode and Forward relay assisted FSO system over k-distribution channel model considering the combine effect of atmospheric turbulence and misalignment error.

In [2] authors have analyzed the behavior of pointing errors and derived the probability density function (pdf) for normalized irradiance due to boresight and jitter. The effect of pointing errors on the performance of the FSO communication system over log normal and gamma-gamma channel models have been studied in [3]-[5]. In [6]-[8] authors, did outage and BER analysis for relay assisted FSO system over strong atmospheric turbulence channel with spatial diversity and misalignment errors for different modulation scheme. [9] Analyzed the performance of free space optical (FSO) system employing subcarrier intensity modulation with differential phase shift keying described by the lognormal atmospheric turbulence and the pointing error effects. In [10] authors, investigate the outage capacity under the influence of atmospheric turbulence and pointing error over gamma distribution model.

The rest of the paper is arranged as follows: Section second discusses the system model and channel model used for proposed work. In section third, the outage probability of relay assisted OWC system is discussed. Section fourth, discusses about bit error rate performance of the system. Section fifth, describes the numerical results and graphical analysis for those results. Finally in section sixth, concluding remarks are highlighted.

## SYSTEM MODEL AND CHANNEL MODEL

### System Model

In this paper we consider a dual-hop FSO system as shown in figure 1. The system model consists of a source (s) and the destination (d) which is communicating with the help of the relay (r). To enable line of sight communication the transceivers are located on the high rise buildings. We assume that due to practical reasons direct communication between source and destination is not feasible which may be either greater distance between source and destination or non-LOS condition. The system employs a full duplex DF relaying protocol. We assume that channel state information (CSI) is present at source, destination and the relay. The laser beam propagates through k-distributed turbulence channel which is corrupted by additive white Guassian noise (AWGN). Further, we consider that channel model is aggregated which takes into account the combined effects of atmospheric turbulence induced fading and misalignment fading (pointing error).

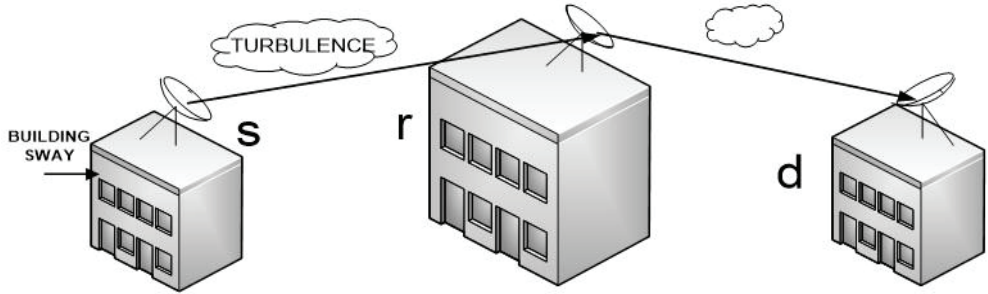

**FIGURE 1.** System Model.

### Channel Model

The k distribution is a widely accepted model for strong atmospheric turbulence. It is considered as a product of gamma distribution and exponential distribution which are two independent models. The probability density function (pdf) of the normalized irradiance 'I' for K distribution is given by,

$$f_I(I) = \frac{2\alpha^{\left(\frac{\alpha+1}{2}\right)}}{\Gamma(\alpha)} I^{\frac{(\alpha-1)}{2}} K_{\alpha-1}(2\sqrt{\alpha I}), \quad I > 0 \tag{1}$$

where channel parameter $\alpha$ is related to the effective no. of discrete scatters, $\Gamma(.)$ is the gamma function [15, eq.(8.310.1)], and $K_v(.)$ is the vth-order modified Bessel function of the second kind [15, eq.(8.432.2)]. By combining pdf for turbulence given in equation (1) and misalignment fading the final pdf of I is given as,

$$f_I(I) = \frac{\alpha g^2}{A_0 \Gamma(\alpha)} G_{1,3}^{3,0}\left(\frac{\alpha}{A_0} I \middle| \begin{array}{c} g^2 \\ -1+g^2, \alpha-1, 0 \end{array}\right) \tag{2}$$

where the ratio between the equivalent beam radius at the receiver and the pointing error displacement standard deviation (jitter) at the receiver is given by $g = \omega_{z_{eq}}/2\sigma_s$. Using the relations $\vartheta = \sqrt{\pi}r/\sqrt{2}\,\omega_z$, $A_0 = \text{erf}(\vartheta)^2$ and $\omega_{z_{eq}}^2 = \omega_z^2 \sqrt{\pi}\,\text{erf}(\vartheta)/2\vartheta \exp(-\vartheta^2)$, the parameter $\omega_{z_{eq}}$ can be calculated where erf(.) is the error function and $\omega_z$ is the beam waist (radius calculated as $e^{-2}$) at distance $z$. The pdf of electrical SNR, $f_\gamma(\gamma)$ can be expressed as,

$$f_\gamma(\gamma) = \frac{\alpha g^2}{A_0 \Gamma(\alpha) \mu^{1/2} \gamma^{1/2}} G_{1,3}^{3,0}\left(\frac{\alpha}{A_0 \sqrt{\mu}} \gamma^{1/2} \middle| \begin{array}{c} g^2 \\ -1+g^2, \alpha-1, 0 \end{array}\right) \tag{3}$$

where instantaneous electrical signal-to-noise ratio (SNR) can be defined as $\gamma = (\eta I)^2/N_0$. The electrical average SNR is defined as $\mu = (\eta E[I]^2)/N_0$. $E[1]=1$ since I is normalized. The CDF of $\gamma$ can be expressed as,

$$F_\gamma(\gamma) = \frac{\alpha g^2}{A_0 \Gamma(\alpha) \mu^{1/2}} \int_0^{\gamma_{th}} \gamma^{\frac{1}{2}-1} G_{1,3}^{3,0}\left(\frac{\alpha}{A_0\sqrt{\mu}}\gamma^{1/2}\Big|_{-1+g^2,\alpha-1,0}^{g^2}\right) d\gamma \tag{4}$$

Let,

$$N = G_{1,3}^{3,0}\left(\frac{\alpha}{A_0\sqrt{\mu}}\gamma^{1/2}\Big|_{-1+g^2,\alpha-1,0}^{g^2}\right) \tag{5}$$

By solving (5), N can be expressed as,

$$N = \frac{2^{\alpha-2}}{2\pi} G_{2,6}^{6,0}\left(\frac{\alpha^2}{16A_0^2\mu}\gamma \Big|_{\frac{-1+g^2}{2},\frac{g^2}{2},\frac{\alpha-2}{2},\frac{\alpha}{2},0,\frac{1}{2}}^{\frac{g^2}{2},\frac{g^2+1}{2}}\right) \tag{6}$$

substitute (6) in (4),

$$F_\gamma(\gamma) = \frac{\alpha g^2 2^{\alpha-2}}{2\pi A_0 \Gamma(\alpha) \mu^{1/2}} \gamma_{th}^{1/2} G_{3,7}^{6,1}\left(\frac{\alpha^2}{16A_0^2\mu}\gamma_{th}\Big|_{\frac{-1+g^2}{2},\frac{g^2}{2},\frac{\alpha-1}{2},\frac{\alpha}{2},0,\frac{1}{2},\frac{-1}{2}}^{\frac{-1}{2},\frac{g^2}{2},\frac{g^2+1}{2}}\right) \tag{7}$$

Equation (7) represents the final CDF of k-distributed turbulence model with misalignment fading (pointing error).

## OUTAGE ANALYSIS

The outage probability is defined as the probability when the instantaneous SNR falls below a specified threshold $\gamma_{th}$ and is given as,

$$Prob(\gamma < \gamma_{th}) = \int_0^{\gamma_{th}} f_\gamma(\gamma) d\gamma \tag{8}$$

$$Prob(\gamma < \gamma_{th}) = F_\gamma(\gamma_{th}) \tag{9}$$

where $\gamma_{th}$ is the threshold SNR. The outage probability $P_{out}$ at the destination is obtained in terms of received SNR given as,

$$P_{out} = P(\gamma_1 < \gamma_{th}) + P(\gamma_2 < \gamma_{th}) - P(\gamma_1 < \gamma_{th})P(\gamma_2 < \gamma_{th}) \tag{10}$$

$$P_{out} = F_{\gamma_1}(\gamma_{th}) + F_{\gamma_2}(\gamma_{th}) - F_{\gamma_1}(\gamma_{th})F_{\gamma_2}(\gamma_{th}) \tag{11}$$

where $\gamma_1$ is the electrical SNR from the source (s) to relay (r) and $\gamma_2$ is the electrical SNR from relay (r) to destination (d). Substituting (7) in (11), the outage probability of *s* to *r* link can be written as,

$$P_{out} = \frac{\alpha_1 g^2 2^{\alpha_1-2}}{2\pi A_0 \Gamma(\alpha_1)\mu_1^{\frac{1}{2}}} \gamma_{th}^{\frac{1}{2}} G_{3,7}^{6,1}\left(\frac{\alpha_1^2}{16A_0^2\mu_1}\gamma_{th}\Big|_{\frac{-1+g^2}{2},\frac{g^2}{2},\frac{\alpha_1-1}{2},\frac{\alpha_1}{2},0,\frac{1}{2},\frac{-1}{2}}^{\frac{-1}{2},\frac{g^2}{2},\frac{g^2+1}{2}}\right) +$$

$$\frac{\alpha_2 g^2 2^{\alpha_2-2}}{2\pi A_0 \Gamma(\alpha_2)\mu_2^{\frac{1}{2}}} \gamma_{th}^{\frac{1}{2}} G_{3,7}^{6,1}\left(\frac{\alpha_2^2}{16A_0^2\mu_2}\gamma_{th}\Big|_{\frac{-1+g^2}{2},\frac{g^2}{2},\frac{\alpha_2-1}{2},\frac{\alpha_2}{2},0,\frac{1}{2},\frac{-1}{2}}^{\frac{-1}{2},\frac{g^2}{2},\frac{g^2+1}{2}}\right) -$$

$$\frac{\alpha_1\alpha_2 g^4 2^{\alpha_1-2} 2^{\alpha_2-2}}{4\pi^2 A_0^2 \Gamma(\alpha_1)\Gamma(\alpha_2)\mu_1^{\frac{1}{2}}\mu_2^{\frac{1}{2}}} \gamma_{th} G_{3,7}^{6,1}\left(\frac{\alpha_1^2}{16A_0^2\mu_1}\gamma_{th}\Big|_{\frac{-1+g^2}{2},\frac{g^2}{2},\frac{\alpha_1-1}{2},\frac{\alpha_1}{2},0,\frac{1}{2},\frac{-1}{2}}^{\frac{-1}{2},\frac{g^2}{2},\frac{g^2+1}{2}}\right) G_{3,7}^{6,1}\left(\frac{\alpha_2^2}{16A_0^2\mu_2}\gamma_{th}\Big|_{\frac{-1+g^2}{2},\frac{g^2}{2},\frac{\alpha_2-1}{2},\frac{\alpha_2}{2},0,\frac{1}{2},\frac{-1}{2}}^{\frac{-1}{2},\frac{g^2}{2},\frac{g^2+1}{2}}\right) \tag{12}$$

Equation (12) represents the closed form expression of outage probability for dual-hop FSO system. Considering the special case with $\alpha_1 = \alpha_2 = \alpha, \mu_1 = \mu_2 = \mu$, we will get closed form expression of outage probability as,

$$P_{out} = \frac{\alpha g^2 2^{\alpha-2}}{\pi A_0 \Gamma(\alpha) \mu^{1/2}} \gamma_{th}^{1/2} G_{3,7}^{6,1}\left(\frac{\alpha^2}{16 A_0^2 \mu}\gamma_{th} \middle| \begin{array}{c} \frac{-1}{2},\frac{g^2}{2},\frac{g^2+1}{2} \\ \frac{-1+g^2}{2},\frac{g^2}{2},\frac{\alpha-1}{2},\frac{\alpha}{2},0,\frac{1}{2},\frac{-1}{2} \end{array}\right) - \left(\frac{\alpha g^2 2^{\alpha-2}}{2\pi A_0 \Gamma(\alpha) \mu^{1/2}}\right)^2 \gamma_{th} \left[G_{3,7}^{6,1}\left(\frac{\alpha^2}{16 A_0^2 \mu}\gamma_{th} \middle| \begin{array}{c} \frac{-1}{2},\frac{g^2}{2},\frac{g^2+1}{2} \\ \frac{-1+g^2}{2},\frac{g^2}{2},\frac{\alpha-1}{2},\frac{\alpha}{2},0,\frac{1}{2},\frac{-1}{2} \end{array}\right)\right]^2$$

(13)

Equation (13) represents the outage probability for single relay assisted FSO systems with pointing error.

## BER ANALYSIS

The average BER can be written as,

$$P_e = \frac{q^P}{2\Gamma(P)} \int_0^\infty \exp(-q\gamma) \gamma^{P-1} F(\gamma) d\gamma \quad (14)$$

where the parameter p and q in (14) account for different modulation schemes. Substituting the value of $F_\gamma(\gamma)$ from equation (12) to equation (14), we get,

$$P_e = \frac{q^P}{2\Gamma(P)} \int_0^\infty \exp(-q\gamma) \gamma^{P-1} \left[\frac{\alpha_1 g^2 2^{\alpha_1-2}}{2\pi A_0 \Gamma(\alpha_1) \mu_1^{\frac{1}{2}}} \gamma_{th}^{\frac{1}{2}} G_{3,7}^{6,1}\left(\frac{\alpha_1^2}{16 A_0^2 \mu_1}\gamma_{th} \middle| \begin{array}{c} \frac{-1}{2},\frac{g^2}{2},\frac{g^2+1}{2} \\ \frac{-1+g^2}{2},\frac{g^2}{2},\frac{\alpha_1-1}{2},\frac{\alpha_1}{2},0,\frac{1}{2},\frac{-1}{2} \end{array}\right) + \frac{\alpha_2 g^2 2^{\alpha_2-2}}{2\pi A_0 \Gamma(\alpha_2) \mu_2^{\frac{1}{2}}} \gamma_{th}^{\frac{1}{2}} G_{3,7}^{6,1}\left(\frac{\alpha_2^2}{16 A_0^2 \mu_2}\gamma_{th} \middle| \begin{array}{c} \frac{-1}{2},\frac{g^2}{2},\frac{g^2+1}{2} \\ \frac{-1+g^2}{2},\frac{g^2}{2},\frac{\alpha_2-1}{2},\frac{\alpha_2}{2},0,\frac{1}{2},\frac{-1}{2} \end{array}\right) - \frac{\alpha_1 \alpha_2 g^4 2^{\alpha_1-2} 2^{\alpha_2-2}}{4\pi^2 A_0^2 \Gamma(\alpha_1)\Gamma(\alpha_2) \mu_1^{\frac{1}{2}} \mu_2^{\frac{1}{2}}} \gamma_{th} G_{3,7}^{6,1}\left(\frac{\alpha_1^2}{16 A_0^2 \mu_1}\gamma_{th} \middle| \begin{array}{c} \frac{-1}{2},\frac{g^2}{2},\frac{g^2+1}{2} \\ \frac{-1+g^2}{2},\frac{g^2}{2},\frac{\alpha_1-1}{2},\frac{\alpha_1}{2},0,\frac{1}{2},\frac{-1}{2} \end{array}\right) G_{3,7}^{6,1}\left(\frac{\alpha_2^2}{16 A_0^2 \mu_2}\gamma_{th} \middle| \begin{array}{c} \frac{-1}{2},\frac{g^2}{2},\frac{g^2+1}{2} \\ \frac{-1+g^2}{2},\frac{g^2}{2},\frac{\alpha_2-1}{2},\frac{\alpha_2}{2},0,\frac{1}{2},\frac{-1}{2} \end{array}\right)\right] d\gamma$$

(15)

Further simplification,

$$P_e = \frac{q^P}{2\Gamma(P)} \left[\frac{\alpha_1 g^2 2^{\alpha_1-2}}{2\pi A_0 \Gamma(\alpha_1) \mu_1^{\frac{1}{2}}} \int_0^\infty \exp(-q\gamma) \gamma_{th}^{(P+\frac{1}{2})-1} G_{3,7}^{6,1}\left(\frac{\alpha_1^2}{16 A_0^2 \mu_1}\gamma_{th} \middle| \begin{array}{c} \frac{-1}{2},\frac{g^2}{2},\frac{g^2+1}{2} \\ \frac{-1+g^2}{2},\frac{g^2}{2},\frac{\alpha_1-1}{2},\frac{\alpha_1}{2},0,\frac{1}{2},\frac{-1}{2} \end{array}\right) d\gamma \right. \\ + \frac{\alpha_2 g^2 2^{\alpha_2-2}}{2\pi A_0 \Gamma(\alpha_2) \mu_2^{\frac{1}{2}}} \int_0^\infty \exp(-q\gamma) \gamma_{th}^{(P+\frac{1}{2})-1} G_{3,7}^{6,1}\left(\frac{\alpha_2^2}{16 A_0^2 \mu_2}\gamma_{th} \middle| \begin{array}{c} \frac{-1}{2},\frac{g^2}{2},\frac{g^2+1}{2} \\ \frac{-1+g^2}{2},\frac{g^2}{2},\frac{\alpha_2-1}{2},\frac{\alpha_2}{2},0,\frac{1}{2},\frac{-1}{2} \end{array}\right) d\gamma \\ - \frac{\alpha_1 \alpha_2 g^4 2^{\alpha_1-2} 2^{\alpha_2-2}}{4\pi^2 A_0^2 \Gamma(\alpha_1)\Gamma(\alpha_2) \mu_1^{\frac{1}{2}} \mu_2^{\frac{1}{2}}} \int_0^\infty \exp(-q\gamma) \gamma_{th}^{(P+1)-1} G_{3,7}^{6,1}\left(\frac{\alpha_1^2}{16 A_0^2 \mu_1}\gamma_{th} \middle| \begin{array}{c} \frac{-1}{2},\frac{g^2}{2},\frac{g^2+1}{2} \\ \frac{-1+g^2}{2},\frac{g^2}{2},\frac{\alpha_1-1}{2},\frac{\alpha_1}{2},0,\frac{1}{2},\frac{-1}{2} \end{array}\right) \\ \left. G_{3,7}^{6,1}\left(\frac{\alpha_2^2}{16 A_0^2 \mu_2}\gamma_{th} \middle| \begin{array}{c} \frac{-1}{2},\frac{g^2}{2},\frac{g^2+1}{2} \\ \frac{-1+g^2}{2},\frac{g^2}{2},\frac{\alpha_2-1}{2},\frac{\alpha_2}{2},0,\frac{1}{2},\frac{-1}{2} \end{array}\right) d\gamma\right]$$

(16)

Let,

$$I_1 = \frac{\alpha_1 g^2 2^{\alpha_1-2}}{2\pi A_0 \Gamma(\alpha_1)\mu_1^{\frac{1}{2}}} \int_0^\infty \exp(-q\gamma)\, \gamma_{th}^{(P+\frac{1}{2})-1} G_{3,7}^{6,1}\left(\frac{\alpha_1{}^2}{16A_0{}^2\mu_1}\gamma_{th} \left|\begin{array}{c} \frac{-1}{2},\frac{g^2}{2},\frac{g^2+1}{2} \\ \frac{-1+g^2}{2},\frac{g^2}{2},\frac{\alpha_1-1}{2},\frac{\alpha_1}{2},0,\frac{1}{2},\frac{-1}{2} \end{array}\right.\right) d\gamma \quad (17)$$

By using [14, eq. (19)],

$$I_1 = \frac{\alpha_1 g^2 2^{\alpha_1-2}}{2\pi A_0 \Gamma(\alpha_1)\mu_1^{\frac{1}{2}}} \int_0^\infty \gamma_{th}^{(P+\frac{1}{2})-1} G_{0,1}^{1,0}\left(q\gamma \left|\begin{array}{c} - \\ 0 \end{array}\right.\right) G_{3,7}^{6,1}\left(\frac{\alpha_1{}^2}{16A_0{}^2\mu_1}\gamma_{th} \left|\begin{array}{c} \frac{-1}{2},\frac{g^2}{2},\frac{g^2+1}{2} \\ \frac{-1+g^2}{2},\frac{g^2}{2},\frac{\alpha_1-1}{2},\frac{\alpha_1}{2},0,\frac{1}{2},\frac{-1}{2} \end{array}\right.\right) d\gamma \quad (18)$$

By solving eq.(18),

$$I_1 = \frac{\alpha_1 g^2 2^{\alpha_1-2}}{2\pi A_0 \Gamma(\alpha_1)\mu_1^{\frac{1}{2}}} q^{-(P+1/2)} G_{4,7}^{6,2}\left(\frac{\alpha_1{}^2}{16A_0{}^2\mu_1 q} \left|\begin{array}{c} \frac{-1}{2},\frac{g^2}{2},\frac{g^2+1}{2},\frac{1}{2}-P \\ \frac{-1+g^2}{2},\frac{g^2}{2},\frac{\alpha_1-1}{2},\frac{\alpha_1}{2},0,\frac{1}{2},\frac{-1}{2} \end{array}\right.\right) \quad (19)$$

Let,

$$I_2 = \frac{\alpha_2 g^2 2^{\alpha_2-2}}{2\pi A_0 \Gamma(\alpha_2)\mu_2^{\frac{1}{2}}} \int_0^\infty \exp(-q\gamma)\, \gamma_{th}^{(P+1/2)-1} G_{3,7}^{6,1}\left(\frac{\alpha_2{}^2}{16A_0{}^2\mu_2}\gamma_{th} \left|\begin{array}{c} \frac{-1}{2},\frac{g^2}{2},\frac{g^2+1}{2} \\ \frac{-1+g^2}{2},\frac{g^2}{2},\frac{\alpha_2-1}{2},\frac{\alpha_2}{2},0,\frac{1}{2},\frac{-1}{2} \end{array}\right.\right) d\gamma \quad (20)$$

By using [14, eq. (44)] and solving,

$$I_2 = \frac{\alpha_2 g^2 2^{\alpha_2-2}}{2\pi A_0 \Gamma(\alpha_2)\mu_2^{\frac{1}{2}}} q^{-(P+1/2)} G_{4,7}^{6,2}\left(\frac{\alpha_2{}^2}{16A_0{}^2\mu_2 q} \left|\begin{array}{c} \frac{-1}{2},\frac{g^2}{2},\frac{g^2+1}{2},\frac{1}{2}-P \\ \frac{-1+g^2}{2},\frac{g^2}{2},\frac{\alpha_2-1}{2},\frac{\alpha_2}{2},0,\frac{1}{2},\frac{-1}{2} \end{array}\right.\right) \quad (21)$$

Let,

$$I_3 = \frac{\alpha_1\alpha_2 g^4 2^{\alpha_1-2} 2^{\alpha_2-2}}{4\pi^2 A_0{}^2 \Gamma(\alpha_1)\Gamma(\alpha_2)\mu_1^{\frac{1}{2}}\mu_2^{\frac{1}{2}}} \int_0^\infty \exp(-q\gamma)\, \gamma_{th}^{(P+1)-1} G_{3,7}^{6,1}\left(\frac{\alpha_1{}^2}{16A_0{}^2\mu_1}\gamma_{th} \left|\begin{array}{c} \frac{-1}{2},\frac{g^2}{2},\frac{g^2+1}{2} \\ \frac{-1+g^2}{2},\frac{g^2}{2},\frac{\alpha_1-1}{2},\frac{\alpha_1}{2},0,\frac{1}{2},\frac{-1}{2} \end{array}\right.\right)$$
$$\times G_{3,7}^{6,1}\left(\frac{\alpha_2{}^2}{16A_0{}^2\mu_2}\gamma_{th} \left|\begin{array}{c} \frac{-1}{2},\frac{g^2}{2},\frac{g^2+1}{2} \\ \frac{-1+g^2}{2},\frac{g^2}{2},\frac{\alpha_2-1}{2},\frac{\alpha_2}{2},0,\frac{1}{2},\frac{-1}{2} \end{array}\right.\right) d\gamma \quad (22)$$

By using [Adamchik,1990, eq. (24)],

$$I_3 = \frac{\alpha_1\alpha_2 g^4 2^{\alpha_1-2} 2^{\alpha_2-2}}{4\pi^2 A_0{}^2 \Gamma(\alpha_1)\Gamma(\alpha_2)\mu_1^{\frac{1}{2}}\mu_2^{\frac{1}{2}}} \int_0^\infty \gamma_{th}^{(P+1)-1} G_{0,1}^{1,0}\left(q\gamma \left|\begin{array}{c} - \\ 0 \end{array}\right.\right)$$
$$\times G_{3,7}^{6,1}\left(\frac{\alpha_1{}^2}{16A_0{}^2\mu_1}\gamma_{th} \left|\begin{array}{c} \frac{-1}{2},\frac{g^2}{2},\frac{g^2+1}{2} \\ \frac{-1+g^2}{2},\frac{g^2}{2},\frac{\alpha_1-1}{2},\frac{\alpha_1}{2},0,\frac{1}{2},\frac{-1}{2} \end{array}\right.\right) G_{3,7}^{6,1}\left(\frac{\alpha_2{}^2}{16A_0{}^2\mu_2}\gamma_{th} \left|\begin{array}{c} \frac{-1}{2},\frac{g^2}{2},\frac{g^2+1}{2} \\ \frac{-1+g^2}{2},\frac{g^2}{2},\frac{\alpha_2-1}{2},\frac{\alpha_2}{2},0,\frac{1}{2},\frac{-1}{2} \end{array}\right.\right) d\gamma \quad (23)$$

By solving eq. (23),

$$I_3 = \frac{\alpha_1\alpha_2 g^4 2^{\alpha_1-2} 2^{\alpha_2-2}}{4\pi^2 A_0{}^2 \Gamma(\alpha_1)\Gamma(\alpha_2)\mu_1^{\frac{1}{2}}\mu_2^{\frac{1}{2}}} q^{-(P+1)}$$
$$\times G_{1,0:3,7:3,7}^{0,1:6,1:6,1}\left(\begin{array}{c}-P \\ -\end{array}\left|\begin{array}{c}\frac{-1}{2},\frac{g^2}{2},\frac{g^2+1}{2} \\ \frac{-1+g^2}{2},\frac{g^2}{2},\frac{\alpha_1-1}{2},\frac{\alpha_1}{2},0,\frac{1}{2},\frac{-1}{2}\end{array}\right|\begin{array}{c}\frac{-1}{2},\frac{g^2}{2},\frac{g^2+1}{2} \\ \frac{-1+g^2}{2},\frac{g^2}{2},\frac{\alpha_2-1}{2},\frac{\alpha_2}{2},0,\frac{1}{2},\frac{-1}{2}\end{array}\right|\left.\frac{\alpha_1{}^2}{16A_0{}^2\mu_1 q},\frac{\alpha_2{}^2}{16A_0{}^2\mu_2 q}\right)$$
$$(24)$$

By using the values of eq. (21), (22) and (24) in eq. (16),

$$P_e = \frac{q^P}{2\Gamma(P)} \left[ \begin{array}{l} q^{-\left(P+\frac{1}{2}\right)} \left[ \begin{array}{l} \frac{\alpha_1 g^2 2^{\alpha_1-2}}{2\pi A_0 \Gamma(\alpha_1)\mu_1^{\frac{1}{2}}} G_{4,7}^{6,2}\left( \frac{\alpha_1^2}{16A_0^2 \mu_1 q} \left| \begin{array}{l} \frac{-1}{2},\frac{g^2}{2},\frac{g^2+1}{2},\frac{1}{2}-P \\ \frac{-1+g^2}{2},\frac{g^2}{2},\frac{\alpha_1-1}{2},\frac{\alpha_1}{2},0,\frac{1}{2},\frac{-1}{2} \end{array} \right. \right) \\ + \frac{\alpha_2 g^2 2^{\alpha_2-2}}{2\pi A_0 \Gamma(\alpha_2)\mu_2^{\frac{1}{2}}} G_{4,7}^{6,2}\left( \frac{\alpha_2^2}{16A_0^2 \mu_2 q} \left| \begin{array}{l} \frac{-1}{2},\frac{g^2}{2},\frac{g^2+1}{2},\frac{1}{2}-P \\ \frac{-1+g^2}{2},\frac{g^2}{2},\frac{\alpha_2-1}{2},\frac{\alpha_2}{2},0,\frac{1}{2},\frac{-1}{2} \end{array} \right. \right) \end{array} \right] \\ - \frac{\alpha_1 \alpha_2 g^4 2^{\alpha_1-2} 2^{\alpha_2-2}}{4\pi^2 A_0^2 \Gamma(\alpha_1)\Gamma(\alpha_2)\mu_1^{\frac{1}{2}}\mu_2^{\frac{1}{2}}} q^{-(P+1)} \\ G_{1,0:3,7:3,7}^{0,1:6,1:6,1}\left( \begin{array}{c} -P \\ - \end{array} \left| \begin{array}{l} \frac{-1}{2},\frac{g^2}{2},\frac{g^2+1}{2} \\ \frac{-1+g^2}{2},\frac{g^2}{2},\frac{\alpha_1-1}{2},\frac{\alpha_1}{2},0,\frac{1}{2},\frac{-1}{2} \end{array} \right| \begin{array}{l} \frac{-1}{2},\frac{g^2}{2},\frac{g^2+1}{2} \\ \frac{-1+g^2}{2},\frac{g^2}{2},\frac{\alpha_2-1}{2},\frac{\alpha_2}{2},0,\frac{1}{2},\frac{-1}{2} \end{array} \left| \frac{\alpha_1^2}{16A_0^2\mu_1 q},\frac{\alpha_2^2}{16A_0^2\mu_2 q} \right. \right) \end{array} \right]$$

(25)

By Considering the special case with $\alpha_1 = \alpha_2 = \alpha, \mu_1 = \mu_2 = \mu, \gamma_1 = \gamma_2 = \gamma$,

$$P_e = \frac{q^P}{2\Gamma(P)} \left[ \begin{array}{l} \frac{\alpha g^2 2^{\alpha-2}}{\pi A_0 \Gamma(\alpha)\mu^{\frac{1}{2}}} q^{-\left(P+\frac{1}{2}\right)} G_{4,7}^{6,2}\left( \frac{\alpha^2}{16A_0^2 \mu q} \left| \begin{array}{l} \frac{-1}{2},\frac{g^2}{2},\frac{g^2+1}{2},\frac{1}{2}-P \\ \frac{-1+g^2}{2},\frac{g^2}{2},\frac{\alpha-1}{2},\frac{\alpha}{2},0,\frac{1}{2},\frac{-1}{2} \end{array} \right. \right) - \frac{\alpha^2 g^4 2^{2(\alpha-2)}}{4\pi^2 A_0^2 [\Gamma(\alpha)]^2 \mu} q^{-(P+1)} \\ G_{1,0:3,7:3,7}^{0,1:6,1:6,1}\left( \begin{array}{c} -P \\ - \end{array} \left| \begin{array}{l} \frac{-1}{2},\frac{g^2}{2},\frac{g^2+1}{2} \\ \frac{-1+g^2}{2},\frac{g^2}{2},\frac{\alpha-1}{2},\frac{\alpha}{2},0,\frac{1}{2},\frac{-1}{2} \end{array} \right| \begin{array}{l} \frac{-1}{2},\frac{g^2}{2},\frac{g^2+1}{2} \\ \frac{-1+g^2}{2},\frac{g^2}{2},\frac{\alpha-1}{2},\frac{\alpha}{2},0,\frac{1}{2},\frac{-1}{2} \end{array} \left| \frac{\alpha^2}{16A_0^2\mu q},\frac{\alpha^2}{16A_0^2\mu q} \right. \right) \end{array} \right]$$

(26)

Equation (26) represents final equation of BER for FSO system over K distribution turbulence channel with pointing errors.

## NUMERICAL RESULTS

An FSO link with turbulence fading parameter $\alpha = 2$ is considered. Fig.2 shows the outage probability in terms of the average SNR (dB) for various values of the g. Where 'g represents' the ratio of equivalent beam radius at the receiver and the pointing error displacement. As the value of g increases outage probability decreases and gives better system performance. Lower value of g represents high misalignment error, as the value of g increases misalignment error decreases. Fig.3 shows the average BER in terms of the average SNR (dB) for various values of the g together with the various modulation schemes. For g = 1.2 and g = 4, system gives better performance for g = 4 which represents lower values of misalignment error. It is clear from the figure that coherent binary phase shift keying (CBPSK) gives better performance than non-coherent binary frequency shift keying (NBFSK). For improving the system performance multiple relay can be used.

| Modulation Scheme | p | q |
|---|---|---|
| Coherent binary phase shift keying | 0.5 | 1 |
| Non-coherent binary frequency shift keying | 1 | 0.5 |

**TABLE 1.** BER Parameters of Binary Modulation.

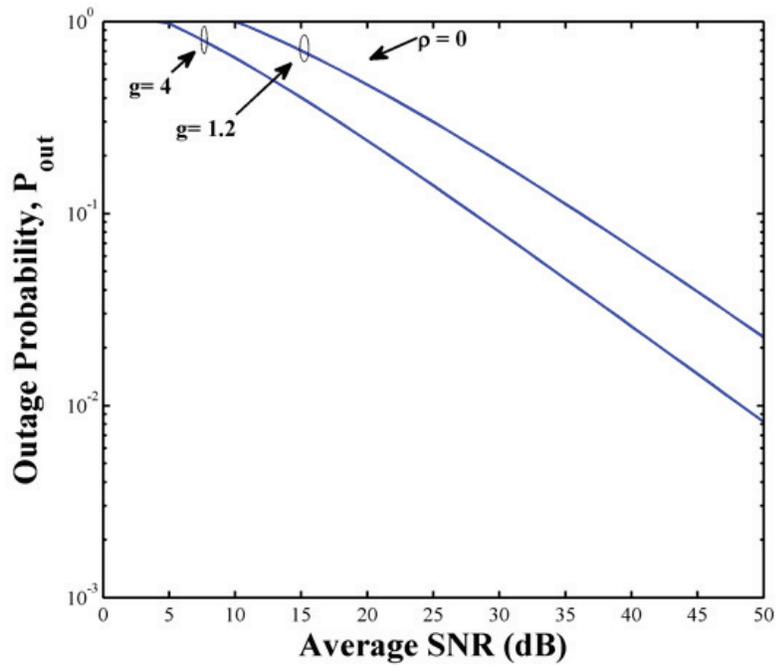

**FIGURE 2.** Outage probability vs. Average SNR (dB) for different value of g for $\alpha = 2$

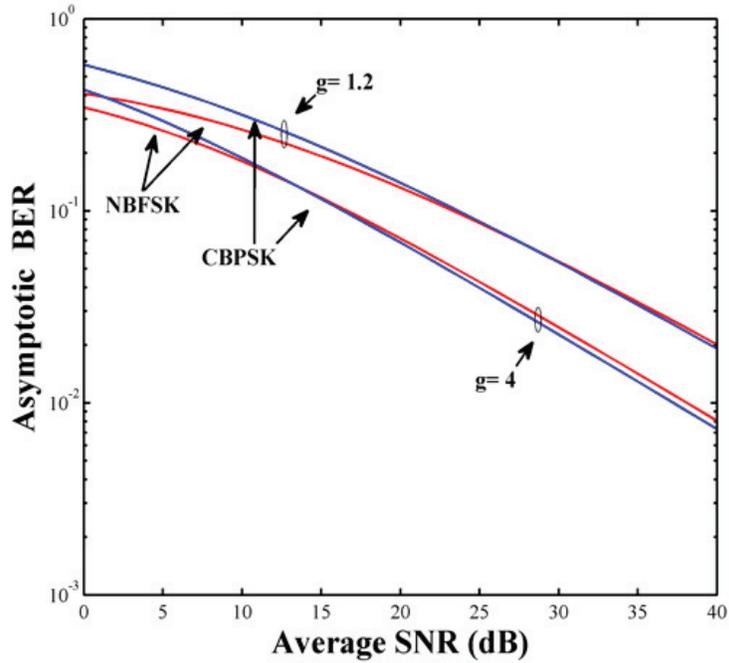

**FIGURE 3.** BER vs. Average SNR (dB) for different modulation schemes.

## CONCLUSION

In the presence of misalignment fading, the outage and BER performance of a FSO system operating over k-distributed turbulence channel, is studied and the closed form expression of outage probability and BER is obtained. We analyze that with an increase in misalignment errors, BER and Outage probability increases and degrade the system performance. Further, we compared the results of different modulation scheme and found out with the BPSK modulation scheme system gives the best performance. It is concluded that for getting excellent system performance even with the strong turbulence conditions analysis can be done for multiple relay system.